\documentclass[12pt]{spieman}  
\usepackage{amsmath,amsfonts,amssymb}
\usepackage{graphicx}
\usepackage{setspace}
\usepackage{tocloft}
\usepackage{lineno}

\title{Wideband bright soliton frequency comb generation at optical telecommunication wavelength in a thin SiN film}

\author[a]{Ali Eshaghian Dorche}
\author[a]{Amir Hossein Hosseinnia}
\author[a]{Ali Asghar Eftekhar}
\author[a,*]{Ali Adibi}
\affil[a]{School of Electrical and Computer Engineering, Georgia Institute of Technology, 778 Atlantic Drive NW, Atlanta, GA, USA 30332}

\cftpagenumbersoff{figure}
\cftpagenumbersoff{table} 
\begin{document} 
\maketitle

\begin{abstract}
Bright-soliton frequency comb generation in a thin-film silicon nitride (SiN) microresonator at optical telecommunication wavelengths is numerically demonstrated using our recently developed approach for dispersion-engineering by virtue of coupling-dispersion between two coupled microresonators. By coupling two identical resonators through an asymmetric Mach-Zehnder structure, sinusoidal splitting of the resonance frequencies can be achieved. This enables the engineering of the dispersion of the resulting modes of the coupled structure. Using this approach, anomalous dispersion of the resonant modes can be achieved at the pumping wavelength (which is selected to be the optical telecommunication wavelength, i.e., 1.55 $\mu m$) to enable Kerr-comb generation. In addition, by utilizing soliton-induced Cherenkov radiation in the coupled-resonator structure, we can increase the bandwidth of the resulting Kerr-comb signal. 
\end{abstract}


{\noindent \footnotesize\textbf{*}Ali Adibi,  \linkable{ali.adibi@ece.gatech.edu} }

\begin{spacing}{2}   


Bright-soliton frequency comb generation at the optical telecommunication wavelength is of great interest due to its application in enhancing the coherent telecommunication capacity through parallel transmission of data streams through the well-defined, low-noise comb lines \cite{marin2017microresonator}, in which, each comb line is used as a carrier in the wavelength-division-multiplexing (WDM) technique. Dissipative Kerr soliton (DKS) features coherent, low-noise comb lines, which can be used as oscillators in the WDM method. The combs associated with DKS inside a microresonator are generated through nonlinear interaction of photons through Kerr nonlinearity of the host material when seeded by the continuous-wave (CW) optical pumping. Thus, by using a coherent low-noise CW source for pumping, simultaneous well-defined, coherent, and low-noise frequency combs can be generated through DKS formation inside a microresonator. The nonlinear interaction is initiated by the modulation instability (MI), followed by cascaded four-wave mixing (FWM) to generate photons at other frequencies. Whispering-gallery-mode (WGM) resonators are widely used for efficient Kerr-comb generation \cite{lin2015dispersion}, especially in CMOS-compatible platform. However, generation of wideband Kerr-comb signals requires precise control of the microresonator dispersion \cite{moss2013new}. Anomalous dispersion of the resonant modes is a pre-requisite for wideband Kerr-comb generation. In addition, low-loss platforms with large nonlinear coefficients are essential for efficient low-power Kerr-comb generation. These requirements enforce limitations on the choice of material platform and device architecture (e.g., geometrical parameters of the microresonators) \cite{moss2013new}.

Among available CMOS-compatible materials, SiN is the preferred platform for Kerr-comb generation due to its wideband transparency window from visible to infrared wavelengths. In addition, the low material loss of SiN enables high quality-factor (Q) microresonators \cite{moss2013new, li2013vertical}. Kerr-comb generation at optical telecommunication wavelengths is demonstrated in high-Q microresonators formed in thick SiN platforms (thickness $>700$ $nm$) to allow for the required dispersion-engineering \cite{okawachi2011octave}. Thick SiN deposition imposes fabrication challenges owing to significant residual stress in the SiN film, which leads to crack formation. Low-pressure chemical vapor deposition (LPCVD) is the preferred method for deposition of SiN due to the superior optical quality (e.g., very low optical loss) of the deposited film. However, it leads to crack formation (caused by significant residual tensile stress) at SiN film thicknesses larger than $450$ $nm$. Thus, more complex fabrication techniques, e.g., optical Damascene process \cite{pfeiffer2016photonic}, generating trenches, and thermal cycling \cite{nam2012patterning} are required to ameliorate crack-formation. Nevertheless, alternative solutions based on thin (low-stress) SiN films (thickness $<450$ $nm$) to mitigate such fabrication complications are of great interest for enabling devices reliable for generation of wideband frequency comb signals that can be integrated with other photonic and/or electronic devices on chip.

To enable Kerr-comb generation at visible and infrared frequencies in thin SiN platforms, concentric thin-SiN microresonators have been recently proposed. However, this approach suffers from significant optical loss (i.e., low Q) or narrowband Kerr-comb \cite{kim2017dispersion, soltani2016enabling}. It is also shown that mode-crossing at the normal dispersion regime in thin SiN films can be helpful in initiating Kerr-comb formation \cite{xue2015normal, liu2014investigation, miller2015tunable}, though suffering from multi-mode characteristic of the microresonator. Mode-crossing is studied in several configurations including a single multi-mode resonator with mode-crossing and coupled resonators with slightly different free-spectral-ranges (FSRs). Nonetheless, the bandwidth of the generated frequency combs in these structures has been limited to 20 nm unless the thick SiN platform (thickness $>700$ $nm$) is used. Another approach is based on dispersion-engineering in a coupled-resonator structure through the use of an asymmetric Mach-Zehnder segment to form sinusoidal coupling between two identical microresonators \cite{dorche2017extending, dorche2018wideband, dorche2018kerr}. Coupled-resonator configuration has eigenmodes (i.e., resonant modes) that deviate from the modes of the individual (identical) microresonators due to mode splitting, which is proportional to the coupling strength. Thus, by engineering the coupling through an asymmetric Mazh-Zehnder segment, a modulated oscillating coupling between the microresonators is achieved. The optimized coupler in the coupled-resonator structure can be used to induce anomalous dispersion at wavelengths where the single resonator has normal dispersion, e.g., $1550$ $nm$, while avoiding thick SiN film.

\begin{figure}
	\centering%
	\includegraphics[trim=0cm 8cm 6cm 0cm,width=12cm]{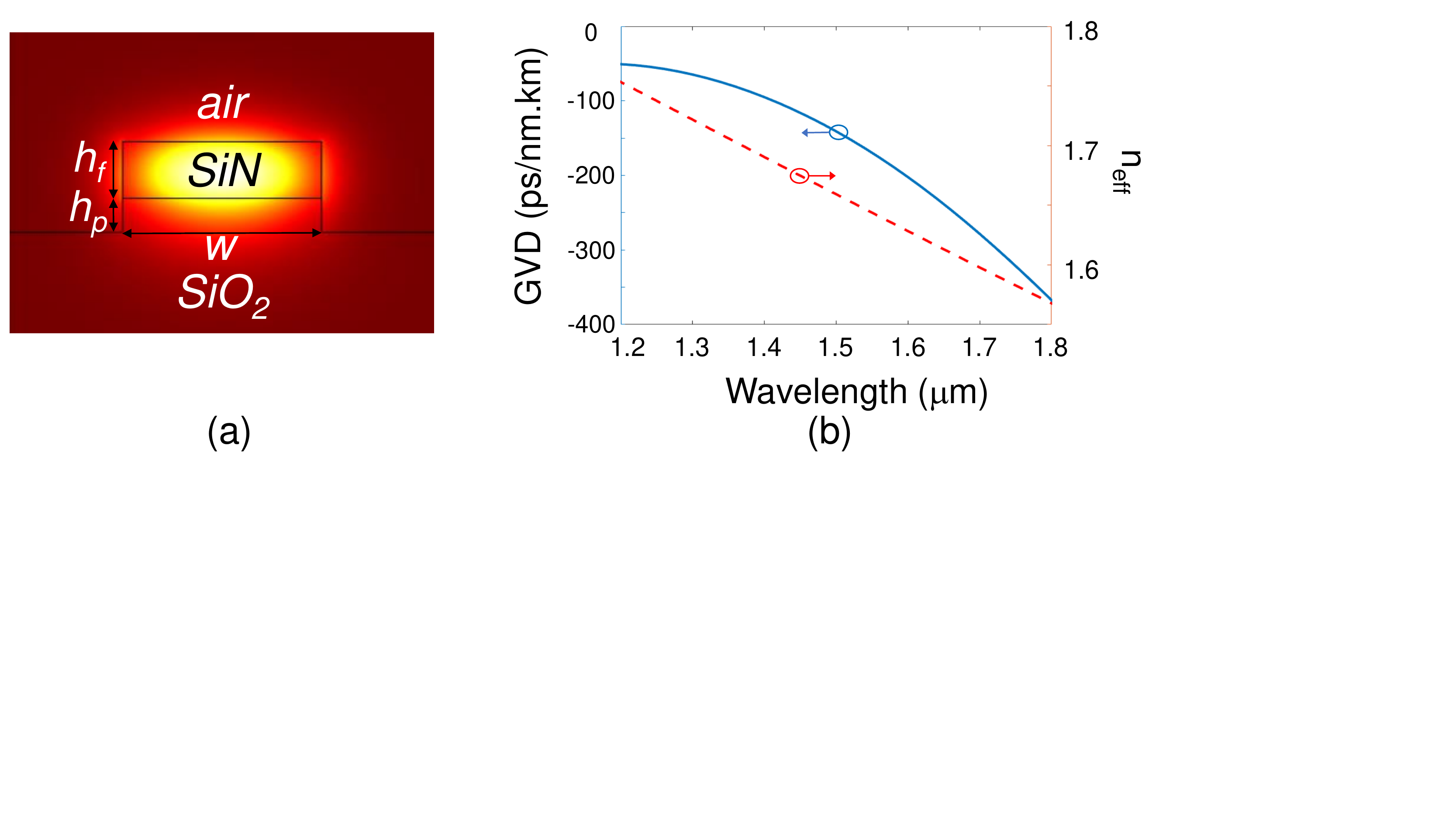}
	\caption{Dispersion engineering of a thin over-etched SiN waveguide: a) mode profile and cross section of the over-etched, air-clad, thin SiN waveguide on a $\mathrm{SiO}_{2}$ substrate. b) Group velocity dispersion parameter of the structure in (a) which demonstrates normal dispersion at $1550$ $nm$ while anomalous dispersion is required for bright-soliton generation. In this simulation, the thickness of the SiN layer, the $\mathrm{SiO}_{2}$ over-etching depth, and the width of the waveguide are $h_f =417$ $nm$, $h_p =250$ $nm$, and $w =1400$ $nm$, respectively.}\label{fig1}
\end{figure}

In this letter, we use our recently demonstrated dispersion-engineering approach based on coupling-dispersion \cite{dorche2017extending} for bright-soliton Kerr-comb generation at optical telecommunication wavelengths. The waveguide structure considered throughout this paper is an over-etched, air-clad, thin $\mathrm{SiN}$-on-$\mathrm{SiO}_{2}$ substrate as depicted in Fig. \ref{fig1}(a). The microresonator formed through bending this waveguiding structure (e.g., in the form of a ring or a racetrack) has, in principle, high Q due to reduced power leakage and increased confinement. In addition, the over-etching can help decreasing the strong normal dispersion. The waveguide geometrical parameters considered throughout the paper are $w =1400$ $nm$, $ h_f =417$ $nm$, $h_p =250$ $nm$. The electric-field profile of the fundamental TE mode (i.e., electric field parallel to the plane of the SiN film) is also shown in Fig. \ref{fig1}(a). This waveguide has normal dispersion at the optical telecommunication wavelengths, as seen from the group-velocity dispersion diagram in Fig. \ref{fig1}(b), obtained by three-dimensional solution of the Maxwell's equation using the finite element method (FEM) implemented in the COMSOL multiphysics environment. Figure \ref{fig1}(b) clearly demonstrates that anomalous dispersion cannot be achieved using a single resonator formed by our waveguiding structure. Thus, we exploit the coupled-resonator structure with an optimized coupling between two identical resonators to ensure anomalous dispersion at the pumping wavelength. Moreover, considering the oscillatory coupling between the microresonators, this structure is expected to take advantage of the soliton-induced Cherenkov radiation to enhance the Kerr-comb bandwidth \cite{brasch2016photonic}.

\begin{figure}
	\centering%
	\includegraphics[trim=3cm 0cm 7cm 3.2cm,width=10.5cm]{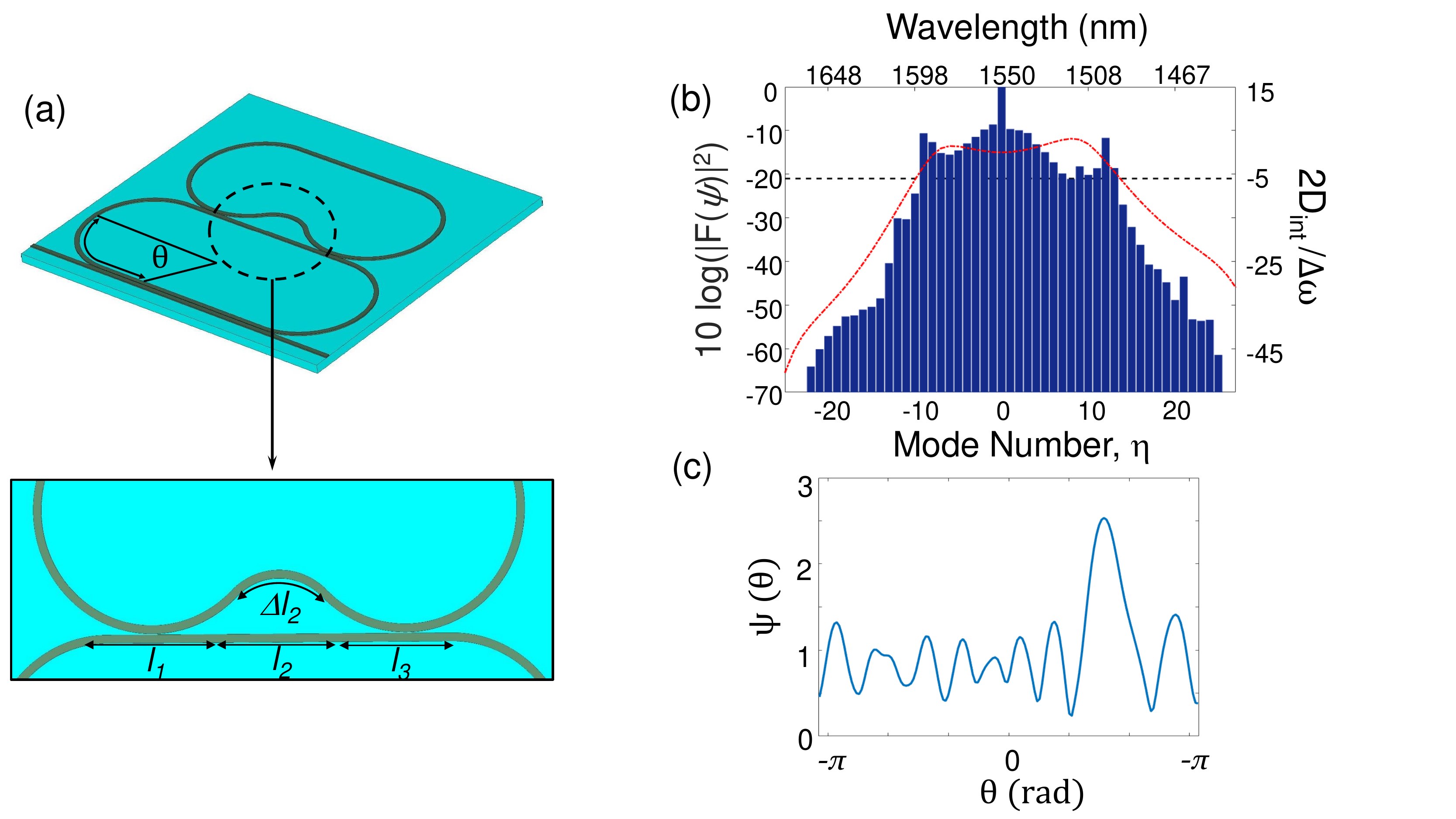}
	\caption{Bright-soliton and soliton-induced Cherenkov radiation in a coupled SiN microresonator structure: a) schematic of the coupled-resonator structure with an asymmetric Mach-Zehnder coupling segment. The total length of the two resonators ($l_{tot}$) are equal, to assure the same FSR. b) Kerr-comb spectrum of the coupled identical microresonators in (a), each having a total perimeter of $80\pi$ $\mu m$, with coupling parameters $l_{1} = 2$ $\mu{m}$, $l_{2} = 1$ $\mu{m}$, $\Delta{l_{2}} = 15$ $\mu{m}$, $l_{3} = 1$ $\mu{m}$, and coupling gap of $150$ $nm$ when excited by a CW pump (at $\lambda=1550$ $nm$) with normalized detuning $\alpha=4$ and normalized power $F^{2}=7$. The peak in the intensity at $\lambda =1594$ $nm$ is due to the soliton-induced Cherenkov radiation. The Kerr-comb is extended from $1447$ $nm$ to $1658$ $nm$ at $-70$ $dB$ window. The dashed-dotted red curve shows the normalized integrated dispersion ($2D_{int}/\Delta\omega$). The regions with convex and concave variation of the normalized dispersion correspond to normal and anomalous dispersion, respectively. The figure clearly shows the intermittent variation between normal and anomalous dispersion in the spectral range between $1447$ $nm$ and $1658$ $nm$. c) Intra-cavity soliton amplitude $|\psi|$, in which the oscillatory tail represents the soliton-induced Cherenkov radiation. The effective angle ($\theta$) at each point on the racetrack resonator is defined as $\theta=2\pi l/l_{tot}$, with $l$ being the length of the corresponding arc on the racetrack resonator}\label{fig2}
\end{figure}

Figure \ref{fig2}(a) illustrates the schematic of the coupled-resonator structure, in which two identical microresonators are coupled through an asymmetric Mach-Zehnder segment. To study the coupling-dispersion-engineering and its applicability for Kerr-comb generation in microresonators with different FSRs, we consider two scenarios where the identical resonators in Fig. \ref{fig2}(a) have total length $l_{tot}=80\pi$ $ \mu m$ and $l_{tot}=400\pi$ $\mu m$. The former (latter) has a larger (smaller) FSR. The coupling between the two resonators in Fig. \ref{fig2}(a) is an oscillatory function of wavelength, which leads to anomalous dispersion with an optimized coupling modulation. We numerically study the dispersion of the resonant modes in the coupled-resonator structure with optimized coupling for both cases to demonstrate the applicability of our approach for bright-soliton Kerr-comb generation. Coupled-mode-theory (CMT) is used to find the resonant modes of the cold coupled-resonator structure \cite{dorche2017extending}, while the exact dispersion of the waveguiding structure (bent to form the resonators) is studied using the FEM implemented in the COMSOL multiphysics environment. In addition, Kerr-comb generation is numerically studied by solving the generalized Lugiato-Lefever equation (LLE), with dispersion parameters of the cold odd-resonant modes, using the split-step Fourier method \cite{chembo2013spatiotemporal, weideman1986split}. During the numerical studies it is supposed that two resonators forming the coupled-resonator structure having the same Q factor, which is reasonable considering very smooth bending curvature in the asymmetric MZI, therefore the two resonators would be synchronized and no additional term is required to be considered in LLE for studying the nonlinear dynamics of the field propagating inside the coupled-cavity structure. The generalized LLE is written as
\begin{equation}
\label{eq:lle}
\dfrac{\mathrm{d}\psi}{\mathrm{d}\tau}=\\
-(1+i\alpha)\psi+(\sum_{n=2}\dfrac{(-i)^{n-1}}{n!}(\dfrac{-2D_{n}}{\Delta{\omega}})\dfrac{\mathrm{d^{n}}\psi}{\mathrm{d}\theta^{n}})+i{\left|\psi\right|}^{2}\psi+F,\\
\end{equation}
\noindent	where $\psi$ is the normalized intra-cavity electric-field amplitude; $\alpha=-2(\Omega_{0}-\omega_{0})/\Delta\omega_{0}$ is the normalized detuning of the pump, $\Omega_{0}$ is the angular frequency of the pumping laser, $F = (2g_0/{\Delta\omega_{0}})^{1/2}F_{0}^{*}$ is the normalized amplitude of the external excitation with $F_0$ being the amplitude of external excitation; and $D_{n}$ is the $n^{th}$ order dispersion parameter in the polynomial (Taylor) expansion of the resonator dispersion around the pumping frequency. Furthermore, $\theta \in [-\pi, \pi]$ is the azimuthal angle, $\tau = \Delta\omega_{0}t/2$ is rescaled time, and $\Delta\omega_{0}$ represents the linewidth of the resonant mode. 

Figure \ref{fig2}(b) shows the normalized spectral power in a coupled microresonator structures formed by two microresonators, each with total length $l_tot=80\pi$ $\mu m$, with coupling parameters $l_{1} = 2$ $\mu{m}$, $l_{2} = 1$ $\mu{m}$, $\Delta{l_{2}} = 15$ $\mu{m}$, $l_{3} = 1$ $\mu{m}$, and coupling gap of $150$ $nm$ when excited by a CW pump (at $\lambda=1550$ $nm$) with normalized detuning of $\alpha=4$ and  input power of $F^2=7$ (which is equal to $126$ $mW$ for loaded $Q = 400$ $K$). The integrated dispersion of the odd-resonance modes is represented as a red line in Fig. \ref{fig2}(b) as well. The phase-matching condition for dispersive wave emission in the form of soliton-induced Cherenkov radiation is shown by the dotted black line. The peaks observed in the Kerr comb spectrum at $\lambda = 1594$ $nm$, and $\lambda = 1500$ $nm$ correspond to the dispersive waves generated through soliton-induced Cherenkov radiation in the coupled-resonator structure with modulated dispersion. It is clear from Fig. \ref{fig2}(b) that the coupled structure with modulated dispersion enables efficient Kerr-comb generation in the optical telecommunication spectrum with spectral coverage from $\lambda = 1447$ $nm$ to $\lambda = 1658$ $nm$ at $-70$ $dB$ window using a thin SiN platform. The $211$ $nm$ bandwidth in this work shows superior performance over that of the recently proposed concentric thin SiN structure \cite{kim2017dispersion} (with bandwidth less than $100$ $nm$ at $-80$ $dB$ spectral window).

Although Fig. \ref{fig2}(b) demonstrates the applicability of coupled microresonators for inducing the anomalous dispersion and efficient DKS frequency-comb generation at the optical telecommunication wavelengths using a thin SiN platform, the number of available comb lines (i.e., resonant modes) in this structure is limited to 47. For massive coherent communications, higher number of available coherent carriers (distinguished frequencies or comb lines) is necessary. This requires resonators with smaller FSR (or larger total length). For this goal, we investigate larger coupled-resonator structure in Fig. \ref{fig2}(a) with total length $l_tot=400\pi$ $\mu m$.

\begin{figure}
	\centering%
	\includegraphics[trim=4cm 2cm 5cm 1.1cm,width=10cm]{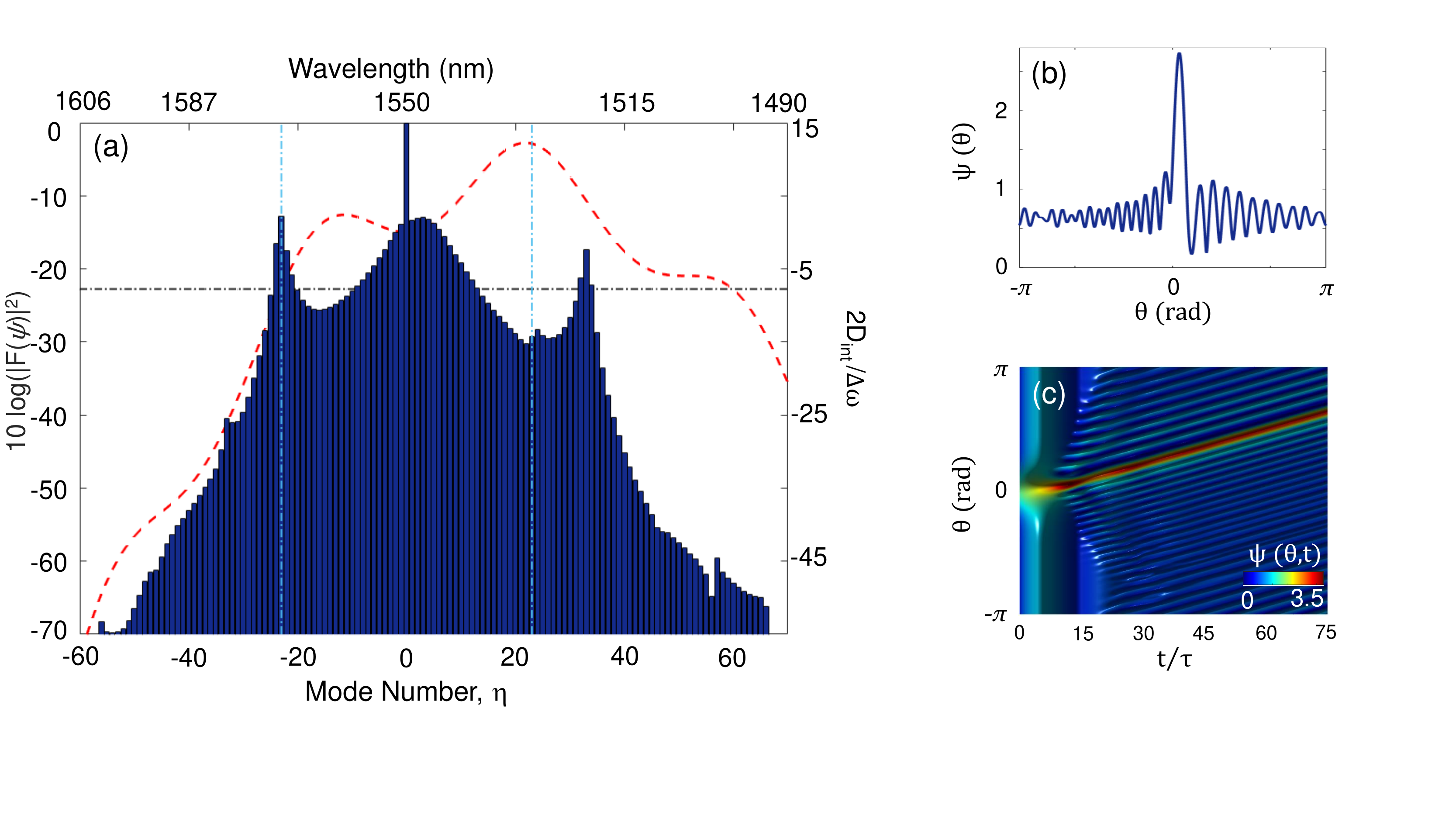}
	\caption{Bright soliton and soliton-induced Cherenkov radiation in a coupled SiN microresonator configuration: a) Kerr-comb spectrum of the coupled microresonator structure in Fig. \ref{fig2}(a), each of them having total length of $400\pi$ $\mu m$ with coupling parameters $l_{1} = 20$ $\mu{m}$, $l_{2} = 1$ $\mu{m}$, $\Delta{l_{2}} = 32$ $\mu{m}$, and $l_{3} = 2$ $\mu{m}$ when excited by a CW pump (at $\lambda=1550$ $nm$) with normalized detuning $\alpha=4$ and normalized power $F^{2}=6$. The peak in intensity at $\lambda =1575$ $nm$ is due to the soliton-induced Cherenkov radiation. The Kerr-comb ranges from $1475$ $nm$ to $1615$ $nm$ at $-70$ $dB$ window. b) Intracavity soliton amplitude whose tail is due to the soliton-induced Cherenkov radiation. The dashed red curve shows the normalized integrated dispersion ($2D_{int}/\Delta\omega$). The regions with convex and concave variation of the normalized dispersion correspond to normal and anomalous dispersion, respectively. The figure clearly shows the intermittent variation between normal and anomalous dispersion in the spectral range between $1447$ $nm$ and $1658$ $nm$. c) Time evolution of the intracavity signal clearly represents the formation of a single bright soliton.}\label{fig3}
\end{figure}

Figure \ref{fig3}(a) shows the normalized Kerr-comb spectrum of the coupled microresonators configuration in Fig. \ref{fig2}(a) with FSR equal to 112 GHz ($l_tot = 400\pi$ $\mu m$), coupling parameters $l_{1} = 20$ $\mu{m}$, $l_{2} = 1$ $\mu{m}$, $\Delta{l_{2}} = 32$ $\mu{m}$, and $l_{3} = 2$ $\mu{m}$ when excited by a CW pump (at $\lambda=1550$ $nm$) with normalized detuning $\alpha=4$ and normalized power $F^{2}=6$. The generated Kerr-comb spans $\lambda = 1493$ $nm$ to $\lambda = 1602$ $nm$ with $122$ comb lines at $-70$ $dB$ window of the pump ($-58$ $dB$ of the peak comb inside the resonator excluding the pump). The normalized integrated dispersion of the odd-resonant eigenmode is also shown as a dashed red line in Fig. \ref{fig3}(a), which represents induced anomalous dispersion in several regions (including the pumping wavelength) separated by regions with the normal dispersion. These cascaded anomalous dispersion regions extend the Kerr-comb bandwidth \cite{dorche2017extending}.

In addition, the higher-order dispersion terms in our coupled resonator structure results in power transfer from the pump (at $\lambda=1550$ $nm$ or $\eta=0$ in Fig. \ref{fig3}(a)) to soliton-induced Cherenkov radiation \cite{brasch2016photonic} (observed as peaks at $\eta=-23$ and $\eta=33$ in Fig. \ref{fig3}(a)). These Cherenkov-radiation peaks act as new oscillators that extend the bandwidth of the Kerr-comb through generating new Kerr-combs that will be eventually phase-locked to the original DKS. Note that the extension of the bandwidth of the Kerr-comb in Fig. \ref{fig3}(a) is primarily caused by engineering the dispersion of the coupled-resonator structure, and it cannot be achieved in a single-resonator thin-SiN structure. The presence of the Cherenkov radiation is also verified by the observation of the soliton recoil (i.e., the spectral shift of the peak of the DKS with respect to pump wavelength ($1550$ $nm$) \cite{yi2017single} in Fig. \ref{fig3}(a)).

It is important to note that the soliton-induced Cherenkov radiation requires phase matching of the generating process; this happens when the integrated dispersion of the cold cavity mode (shown by the dashed red line in Fig. \ref{fig3}(a)) is compensated by the combination of the intracavity energy and the pump detuning, which is calculated using the formulation in Ref. \cite{dorche2017extending}, and it is shown by the horizontal black dash-dotted line in Fig. \ref{fig3}(a). The Cherenkov radiation is expected at the intersect of the two curves (at $\eta=-23$ and $\eta=58$ in Fig. \ref{fig3}(a)). While the Cherenkov radiation at $\eta=-23$ is clearly observed in Fig. \ref{fig3}(a), that at $\eta=58$ undergoes a red-shift to $\eta=33$. To explain the reason for this shift, we have shown the spatial distribution of the intracavity signal as a function of angle ($\theta$) in Fig. \ref{fig3}(b). The oscillatory behavior on the two tails of the soliton are due to the two Cherenkov-radiation signals. Figure \ref{fig3}(b) also shows the interference of these two signals \cite{pfeiffer2017octave}. As a result, the governing equations for phase matching will change, resulting in an additional phase shift, which can cause the spectral shift of the Cherenkov radiation. The extent of this spectral shift due to phase perturbation (caused by the non-negligible amplitude of one Cherenkov radiation at the location of the peak of the other Cherenkov radiation as the perturbing agent) depends on the strength of each Cherenkov radiation amplitude at the location of the other Cherenkov radiation and the slope of the total integrated dispersion of the cold cavity. Both of these parameters are considerably larger for the Cherenkov radiation at $\eta=-23$ as compared to that at $\eta=58$. As a result, the spectral shift is negligible for the former and is considerable for the latter. The time evolution of the intracavity signal is illustrated in Fig. \ref{fig3}(c), clearly demonstrating single bright-soliton formation in the coupled-resonator structure, which confirms that the simulations have been run for enough time to ensure the steady-state nature of the observed spectrum in Fig. \ref{fig3}(a).

\begin{figure}
	\centering%
	\includegraphics[trim=0cm 2cm 7cm 0cm,width=10cm]{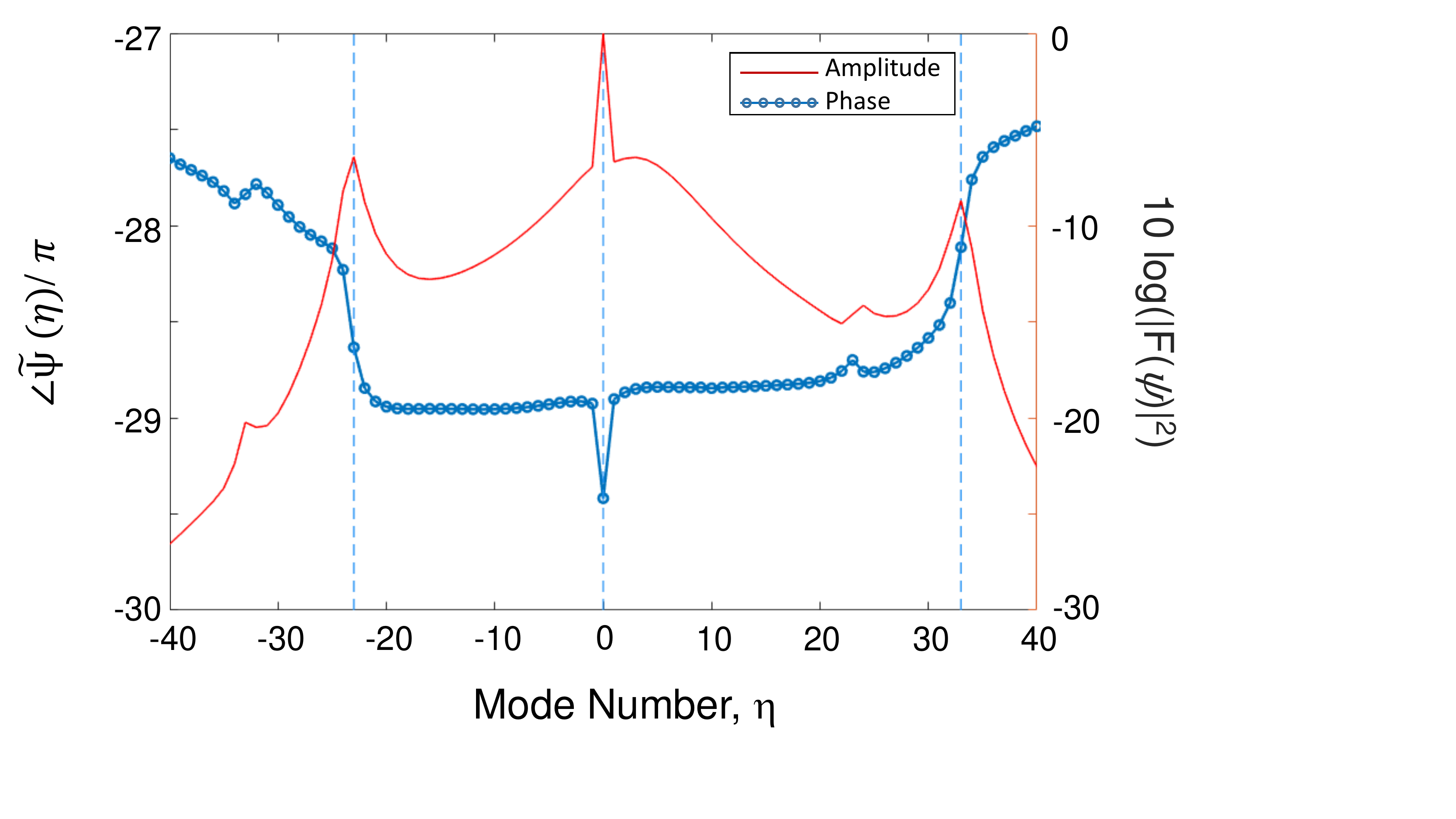}
	\caption{The amplitude and spectral phase of the Kerr-comb generated by the fundamental soliton and the dispersive waves in a coupled-resonator structure of Fig. \ref{fig3}. The sharp phase transitions at $\eta=-23$, and $\eta=33$ match the peaks of the Kerr-comb amplitude spectrum; thus, resonant power transfer from the pump to the dispersive waves. The other small phase shifts at $\eta=-33$, and $23$ are due to the phase conjugate of the dispersive waves.}\label{fig4}
\end{figure}

In order to study the nature of the second emission in Fig. \ref{fig3}(a), we conduct several numerical studies by artificially inducing large optical loss at the peak of the Cherenkov radiation at $\eta=-23$ and its surrounding comb lines. Our simulation showed a blue-shift in the spectral location of the second Cherenkov radiation towards the unperturbed phase-matching condition (from $\eta=33$ towards $\eta=58$). This clearly verifies our explanation of the spectral shift of the second Cherenkov radiation peak. In addition, Figure \ref{fig4} shows the amplitude and phase of the Kerr-comb signal as a function of the mode number ($\eta$) for the structure studied in Fig. \ref{fig3}. Figure \ref{fig4} clearly shows abrupt phase change at the three peaks of the amplitude (i.e., at $\eta=-23$, $\eta=0$, $\eta=33$). The simultaneous presence of an amplitude peak and a phase jump at each frequency shows the presence of a resonance signal (with spectral representation in the form of $A(\omega)= A_{0}/[1+j(\frac{\omega-\omega_{0}}{\Delta\omega})]$ with $A_0$ being the amplitude, $\omega_0$ being the center frequency and $\Delta\omega$ being the bandwidth). Thus, Fig. \ref{fig4} proves that the steady-state signal in our coupled-resonator structure is due to the DKS (at $\eta=0$) and two Cherenkov-radiation at $\eta=-23$ and $\eta=33$. The last one is indeed the spectrally red-shifted Cherenkov radiation observed in Fig. \ref{fig3}(a).

In conclusion, we demonstrated here a new approach for dispersion-engineering in a coupled-resonator structure formed in a thin SiN film. By controlling the amount of the splitting of the modes of the coupled resonators through changing the coupling strength, the desired dispersion of the eigenmodes of the coupled structure can be achieved. We showed that the oscillatory perturbation of the eigenmodes in the coupled-resonator structure can induce anomalous dispersion and soliton-induced Cherenkov radiation. The induced anomalous dispersion enables bright-soliton formation in the thin SiN platform, while the soliton-induced Cherenkov radiation increases the bandwidth of the generated Kerr-comb signal. Using these two unique properties, we numerically showed the feasibility of the formation of bright soliton and wideband (bandwidth $>200$ $nm$) Kerr-combs at the optical telecommunication wavelengths (around $1550$ $nm$), which to the best of our knowledge has the largest bandwidth demonstrated to date for Kerr-combs in the thin SiN platform. Such structure can be effectively used for high-capacity optical telecommunications by providing a large number of low-noise coherent comb lines, which can be used for WDM communication.


\acknowledgments 
This work is supported by Air Force Office of Scientific Research (FA9550-15-1-0342, Dr. Gernot Pomrenke).


\bibliography{sample}   
\bibliographystyle{spiejour}   


\listoffigures

\end{spacing}
\end{document}